# Anomalous resistivity and the origin of a heavy mass in the two-band Hubbard model with one narrow band

M. Yu. Kagan[1], V.V. Val'kov[2]

[1]*P.L. Kapitsa Institute for Physical Problems, Kosygina st. 2, 119334 Moscow, Russia*

+7(499)137-79-85

+7(495)651-21-25

kagan@kapitza.ras.ru

[2]*Kirenskii Institute of Physics, Akademgorodok 50, building 38,*

*660036 Krasnoyarsk, Russia*

vvv@iph.krasn.ru

***Dedicated to the memory of Prof. A.G. Aronov***

**Abstract.** We search for marginal Fermi-liquid behavior [1] in the two-band Hubbard model with one narrow band. We consider the limit of low electron densities in the bands and strong intraband and interband Hubbard interactions. We analyze the influence of electron polaron effect [2] and other mechanisms of mass-enhancement (related to momentum dependence of the self-energies) on effective mass and scattering times of light and heavy components in the clean case (electron – electron scattering and no impurities). We find the tendency towards phase-separation (towards negative partial compressibility of heavy particles) in a 3D case for large mismatch between the densities of heavy and light bands in a strong coupling limit. We also observe that for low temperatures and equal densities the resistivity in a homogeneous state $R(T) \sim T^2$ – behaves in a Fermi-liquid fashion both in 3D and 2D cases. For temperatures higher then effective bandwidth for heavy electrons $T > W_h^*$ the coherent behavior of heavy component is totally destroyed. The heavy particles move diffusively in the surrounding of light particles. At the same time the light particles scatter on the heavy ones as if on immobile (static) impurities. In this regime the heavy component is marginal, while the light one is not. The resistivity goes on saturation for $T > W_h^*$ in the 3D case. In 2D the resistivity has a maximum and localization tail due to weak – localization corrections of Altshuler – Aronov type [3]. Such behavior of resistivity in 3D could be relevant for some uranium-based heavy-fermion compounds like $UNi_2Al_3$ and in 2D for some other mixed-valence compounds possibly including the layered manganites. We also consider briefly the superconductive (SC) instability in the model. The leading instability is towards p-wave pairing and is governed by enhanced Kohn – Luttinger [4] mechanism of SC at low electron density. The critical temperature corresponds to the pairing of heavy electrons via polarization of the light ones in 2D.

# 1 Introduction

The physics of uranium-based heavy-fermion compounds and the origin of a heavy mass $m_h^* \sim 200 m_e$ for f-electrons in them is possibly very different (see [2]) from the physics of cerium-based heavy-fermions, where the Kondo – effect (or more generally the physics of Kondo-lattice model) is dominant [5,6]. The point is that uranium-based heavy-fermions are usually in the mixed- valence limit [7] with strong hybridization between heavy (f-electrons or f-d electrons) and light (s-p electrons) components. On the level of two-particle hybridization interband Hubbard interaction leads to an additional enhancement of the heavy electrons mass due to electron polaron effect (EPE). Physically EPE is connected with a nonadiabatical part of the many-body wave function describing a heavy electron and a cloud of virtual electron – hole pairs of light particles. These pairs are mixed with the wave function of the heavy electrons but do not follow it when a heavy electrons tunnels from one elementary cell to a neighboring one. It is shown in [2] that in the unitary limit of the strong Hubbard interaction between heavy and light electrons an effective heavy mass could reach the value $m_h^*/m_L \sim (m_h/m_L)^2$ and if we start from the ratio $m_h/m_L \sim 10$ between bare masses of heavy and light electrons, on the level of LDA-approximation, for example, we could finish with an effective value $m_h^* \sim 100\ m_L$, which is typical for uranium-based heavy-fermion compounds.

The similar effect could be described also with the help of the strong one particle hybridization between heavy and light bands [2].

A natural question arises: whether the two-band Hubbard model with one narrow band is a simple toy-model to observe non-Fermi liquid behavior and in particular a well-known marginal Fermi liquid behavior [1]. Remind that in marginal Fermi liquid (MFL) theory the quasiparticles are strongly damped (Im$\varepsilon$ ~ Re$\varepsilon$ ~ $T$). The strong damping $\gamma \sim T$ of the quasiparticles (instead of a standard damping for Landau Fermi liquid picture $\gamma \sim T^2/\varepsilon_F$) could explain, according to [1] a lot of experiments in HTSC-compounds including a linear resistivity $R(T) \sim T$ for $T > T_C$ at optimal doping concentrations. The MFL picture was also proposed to describe the properties of UPt$_3$ doped by Pd including the specific heat measurements [8].

In the present paper we evaluate the damping and transport times for heavy and light electrons. We verify these times on marginality and find that for low temperatures $T < W_h^*$ ($W_h^*$ is the effective bandwidth for heavy electrons) and equal densities of heavy and light bands in a homogeneous state we have a standard behavior for Landau Fermi liquid with a resistivity $R(T) \sim T^2$ for the case of electron – electron scattering both in 3D and 2D. For higher temperatures $T > W_h^*$ ($W_h^* \sim 50$ K for $m_h^* \sim 200 m_e$) the heavy band is totally destroyed and heavy particles move diffusively in the surrounding of light particles while the light particles scatter on the heavy ones as if on immobile (static) impurities. For these temperatures the heavy component is marginal, while the light one is not. We try to find a marginal behavior of the light component taking into account also weak localization corrections of Altshuler – Aronov type [3] for scattering time of light electrons. We do not get marginal behavior of light component, but we get a very interesting anomalous resistivity characteristics especially in a 2D case, where for $T \sim W_h^*$ resistivity has a maximum and a localization tail at higher temperatures [9]. In 3D the resistivity goes to saturation for $T > W_h^*$. Such resistivity characteristics could possibly describe some 3D uranium-based heavy-fermion compounds like $UNi_2Al_3$ and some other mixed-valence systems. In 2D the behavior of resistivity possibly has some relation to layered manganites where we deal with two degenerate ($e_g$) conducting orbitals (bands) of d-electrons of Mn. However for manganites an alternative explanation is possible [10]. According to it the resistivity is governed by electron tunneling from one metallic FM polaron to a neighboring one via an insulating AFM or PM-barrier in the regime of a nanoscale phase separation in electronic subsystem. It will be interesting to compare these two mechanisms for resistivity in layered manganites in more detail.

We also consider other mechanisms of heavy mass enhancement different from EPE and find a very pronounced effect in 3D connected with momentum dependence of the self-energy of heavy electrons due to "heavy - light" interaction. In a strong coupling limit this effect could provide even larger ratios of $m_h^*/m_h$ than EPE. It leads to negative compressibility of heavy particles and thus reveals the tendency towards phase-separation or at least charge redistribution between the bands for a large density mismatch $n_h \gg n_L$ in qualitative agreement with the results of [11].

In the final section of the paper we study the leading SC instability which arises in the two-band model in a 2D case. The leading instability at low density is proved to be towards triplet p-wave pairing. It describes the pairing of heavy electrons via polarization of light electrons [12,13] in the framework of the enhanced Kohn – Luttinger [4] mechanism of SC and provides rather realistic critical temperatures in a 2D or layered case, especially for the situation of the geometrically separated bands belonging to neighboring layers.

## 2 The two-band Hubbard model with one narrow band

The Hamiltonian for the two-band Hubbard model reads:

$$\hat{H}' = -t_h \sum_{<ij>\sigma} a_{i\sigma}^+ a_{j\sigma} - t_L \sum_{<ij>\sigma} b_{i\sigma}^+ b_{j\sigma} - \varepsilon_0 \sum_{i\sigma} n_{i\sigma}^h - \mu \sum_{i\sigma} (n_{i\sigma}^L + n_{i\sigma}^h) + \\ + U_{hh} \sum_i n_{ih}^\uparrow n_{ih}^\downarrow + U_{LL} \sum_i n_{iL}^\uparrow n_{iL}^\downarrow + \frac{U_{hL}}{2} \sum_i n_{iL} n_{ih}$$ , (1)

where $U_{hh}$ and $U_{LL}$ are intraband Hubbard interactions for heavy and light electrons respectively, $U_{hL}$ – is interband Hubbard interaction between heavy and light electrons, $t_h$ and $t_L$ are transfer integrals for heavy and light electrons, $n_{ih}^\sigma = a_{i\sigma}^+ a_{i\sigma}, n_{iL}^\sigma = b_{i\sigma}^+ b_{i\sigma}$ – are the densities of heavy and light electrons on site $i$ with spin – projection $\sigma$, $\mu$ – is chemical potential. Note that $-\varepsilon_0$ is the center of gravity of heavy band, $\Delta$ – the difference between the bottoms of the bands is given by $\Delta = -\varepsilon_0 + \frac{(W_L - W_h)}{2} = (E_{min}^h - E_{min}^L)$. After Fourier-transformation we get:

$$\hat{H}' = \sum_{p\sigma} \varepsilon_h(p) a_{p\sigma}^+ a_{p\sigma} + \sum_{p\sigma} \varepsilon_L(p) b_{p\sigma}^+ b_{p\sigma} + U_{hh} \sum_{pp'q} a_{p\uparrow}^+ a_{p'\downarrow}^+ a_{p-q\downarrow} a_{p'+q\uparrow} + \\ + U_{LL} \sum_{pp'q} b_{p\uparrow}^+ b_{p'\downarrow}^+ b_{p-q\downarrow} b_{p'+q\uparrow} + \frac{U_{hL}}{2} \sum_{\substack{pp'q \\ \sigma\sigma'}} a_{p\sigma}^+ (b_{p'\sigma'}^+ b_{p-q\sigma'}) a_{p'+q\sigma}$$ , (2)

where in $D$ – dimensions for the hypercubic lattice $\varepsilon_h(p) = -2t_h \sum_{a=1}^{D} \cos(p_a d) - \varepsilon_0 - \mu$, $\varepsilon_L(p) = -2t_L \sum_{a=1}^{D} \cos(p_a d) - \mu$ – are the quasiparticle energies for heavy and light bands (see Fig. 1), $p_a = \{p_x, p_y, …\}$

– are Cartesian projections of the momentum. For low densities of heavy and light components $n_{tot}d^D = (n_h + n_L)d^D \ll 1$ the quasiparticle spectra read:

$$\varepsilon_h(p) = -\frac{W_h}{2} + t_h(p^2 d^2) - \varepsilon_0 - \mu;$$
$$\varepsilon_L(p) = -\frac{W_L}{2} + t_L(p^2 d^2) - \mu, \qquad (3)$$

where $W_h = 4D\, t_h$ and $W_L = 4D\, t_L$ – are the bandwidths of heavy and light electrons for the $D$ – dimensional hypercubic lattice, $d$ – is intersite distance. Hence introducing the bare masses of heavy and light component:

$$m_h = \frac{1}{2t_h d^2};\ m_L = \frac{1}{2t_L d^2} \qquad (4)$$

and Fermi energies:

$$\varepsilon_{Fh} = \frac{p_{Fh}^2}{2m_h} = \frac{W_h}{2} + \mu + \varepsilon_0;\ \varepsilon_{FL} = \frac{W_L}{2} + \mu, \qquad (5)$$

we finally get for the quasiparticle spectra for $T \to 0$:

$$\varepsilon_h(p) = \frac{p^2}{2m_h} - \varepsilon_{Fh};\ \varepsilon_L(p) = \frac{p^2}{2m_L} - \varepsilon_{FL}. \qquad (6)$$

In deriving (4)-(6) we implicitly assume that the difference between the bottom of the bands $\Delta$ on Fig.1 is not too large, so parabolic approximation for the spectra of both bands is still valid. Note that there is no one-particle hybridization in the Hamiltonians (1,2) but there is a strong two-particle hybridization $\frac{U_{hL}}{2}\sum_i n_i^h n_i^L$.

We assume that $m_h \gg m_L$ and thus

$$W_h / W_L = m_L/m_h \ll 1. \qquad (7)$$

We also assume that $U_{hh} \sim U_{LL} \sim U_{hL} \gg W_L \gg W_h$ – strong-coupling situation ($U_{hL}$ is large because in reality light particles experience strong scattering on the heavy ones as if on a quasiresonance level). Finally we consider the most simple case when densities of the bands are of the same order: $n_h \sim n_L$ (note that in 3D $n = p_F^3/3\pi^2$ while in 2D $n = p_F^2/2\pi$).

# 3 The Kanamori T- matrix approximation

According to renormalization scheme of Kanamori the strong Hubbard interactions [14] in case of low electron density (practically empty lattice) should be described in terms of the corresponding vacuum T-matrices (see Fig. 2).

In the 3D case the solution of the corresponding Bethe – Salpeter integral equations in vacuum yields for T-matrices (see [12-14]):

$$T_{hh} = \frac{U_{hh}d^3}{(1 - U_{hh}d^3 K_{hh}^{vac}(0,0))} = \frac{U_{hh}d^3}{\left(1 + \frac{U_{hh}}{8\pi t_h}\right)};$$

$$T_{hL} = \frac{U_{hL}d^3}{\left(1 + \frac{U_{hL}}{8\pi t_{hL}^*}\right)}; \quad T_{LL} = \frac{U_{LL}d^3}{\left(1 + \frac{U_{LL}}{8\pi t_L}\right)},$$

(8)

where $K_{hh}^{vac}(0,0) = -\int \frac{d^3\vec{p}}{(2\pi)^3} \frac{m_h}{p^2}$ - is a Cooper loop for heavy particles in vacuum (a product of two vacuum Green functions of heavy particles in a Cooper channel for total frequency and total momentum equal to zero), $m_{hL}^* = \frac{1}{2t_{hL}^* d^2} = \frac{m_h m_L}{(m_h + m_L)} \approx m_L$ for $m_h \gg m_L$ – is an effective mass for the T-matrix $T_{hL}$ (for scattering of light electrons on heavy ones) and accordingly $t_{hL}^* \approx t_L$ is an effective transfer integral; $Ud^3$ – plays the role of zeroth Fourier component in 3D. As a result for $U_{hh} \sim U_{LL} \sim U_{hL} \gg W_L \gg W_h$:

$$T_{hh} = 8\pi t_h d^3; T_{hL} \approx T_{LL} \approx 8\pi t_L d^3 \qquad (9)$$

The s-wave scattering length for the Hubbard model [12] is defined as $a = \frac{mT}{4\pi} = \frac{T}{8\pi t d^2}$ and thus:

$$a_{hh} = a_{hL} = a_{LL} = d \qquad (10)$$

in a strong-coupling case.

Correspondingly the gas parameter of Galitskii $f_0 = 2ap_F/\pi$ [15,16] for the case of equal densities of heavy and light bands $n_L = n_h$ reads:

$$f_0 = (f_0^L = 2dp_{FL}/\pi) = (f_0^h = 2dp_{Fh}/\pi) = 2dp_F/\pi. \qquad (11)$$

(it is convenient to include the factor $2/\pi$ in the definition of the gas-parameter in 3D). In the 2D case for strong Hubbard interactions and low densities with logarithmic accuracy the vacuum T-matrices read for $n_L = n_h$ [12,13]:

$$T_{hh} = \frac{U_{hh}d^2}{\left(1+\frac{U_{hh}}{8\pi t_h}\int_{\sim p_F^2}^{\sim 1/d^2}\frac{dp^2}{p^2}\right)} = \frac{U_{hh}d^2}{\left(1+\frac{U_{hh}}{8\pi t_h}\ln\frac{1}{p_F^2 d^2}\right)};$$

$$T_{LL} = \frac{U_{LL}d^2}{\left(1+\frac{U_{LL}}{8\pi t_L}\ln\frac{1}{p_F^2 d^2}\right)};\quad T_{hL} \approx \frac{U_{hL}d^2}{\left(1+\frac{U_{hL}}{8\pi t_L}\ln\frac{1}{p_F^2 d^2}\right)},$$
(12)

where $Ud^2$ plays the role of zeroth Fourier component of the Hubbard potential in 2D. As a result in a strong coupling case the 2D gas parameter of Bloom [17] for equal densities $n_L = n_h$ reads:

$$f_0 = f_{0L} = f_{0h} = \frac{1}{2\ln\frac{1}{p_F d}}. \quad (13)$$

## 4 Evaluation of the self-energies of heavy and light bands

Let us evaluate the imaginary part of the self-energies Im$\Sigma$ in a two-band Hubbard model considering a clean case (no impurities) and taking into account only electron – electron scattering. It is important fro evaluation of the scattering times for heavy and light electrons and further calculation of the resistivity $R(T)$.
In the two-band model (see Fig. 3):

$$\Sigma_h = \Sigma_{hh} + \Sigma_{hL} \text{ and } \Sigma_L = \Sigma_{LL} + \Sigma_{Lh}. \quad (14)$$

The full T-matrices in substance which enter in the diagrams for $\Sigma_{hh}$ in Fig. 3 have the form in 3D case:

$$T_{hh}(\Omega,\vec{p}) = \frac{U_{hh}d^3}{\left(1-U_{hh}d^3 K_{hh}(\Omega,\vec{p})\right)}, \quad (15)$$

where:

$$K_{hh}(\Omega, \vec{p}) = \int \frac{d^3 \vec{p}'}{(2\pi)^3} \frac{(1 - n_h^F(\varepsilon_{p'+p}) - n_h^F(\varepsilon_{-p'}))}{(\Omega - \varepsilon_h(p'+p) - \varepsilon_h(-p') + i0)} \quad - (16)$$

is a Cooper loop in substance (a product of the two Green-functions in the Cooper channel), $n_h^F(\varepsilon)$ is the Fermi – Dirac distribution function for heavy particles, and analogously for the full T-matrices $T_{hL}$, $T_{Lh}$ and $T_{LL}$ and Cooper loops $K_{hL}$, $K_{Lh}$ and $K_{LL}$

If we expand the T-matrix for heavy particles in first two orders in gas-parameter, than according to Galitskii [15] we get:

$$T_{hh}(\Omega, \vec{p}) = \frac{4\pi a_h}{m_h} + \left(\frac{4\pi a_h}{m_h}\right)^2 (K_{hh} - K_{hh}^{vac}) + o\left[\left(\frac{4\pi a_h}{m_h}\right)^3 (K_{hh} - K_{hh}^{vac})^2\right], \quad (17)$$

where

$$\frac{4\pi a_h}{m_h} \approx \frac{U_{hh} d^3}{(1 - U_{hh} d^3 K_{hh}^{vac})} \quad (18)$$

and coincides with Kanamori approximation for the vacuum T-matrix $K_{hh}^{vac}(\Omega, \vec{p}) = \int \frac{d^3 \vec{p}'/(2\pi)^3}{(\Omega - \frac{(\vec{p}' + \vec{p})^2}{2m_h} - \frac{p'^2}{2m_h})}$ is the Cooper loop in vacuum (rigorously speaking the scattering length is defined via $K_{hh}^{vac}(0,0)$ but the difference between $K_{hh}^{vac}(\Omega, \vec{p})$ and $K_{hh}^{vac}(0,0)$ is proportional to the gas-parameter $a_h p_{Fh}$ and is small). $K_{hh}$ in (17) is full Cooper loop (cooperon) in substance for heavy particles given by (16). If we consider the low densities and the energies close to $\varepsilon_F$ we can show that the terms which we neglect in $T_{hh}$ are small with respect to the gas parameter $\frac{4\pi a_h}{m_h}(K_{hh} - K_{hh}^{vac}) \sim a_h p_{Fh}$. The self-energy of heavy particles $\Sigma_{hh}$ in the first two orders of the gas-parameteris given by:

$$\Sigma_{hh}(p) = \sum_k T_{hh}(k+p) G_h(k) \approx \frac{4\pi a_h}{m_h} \sum_k G_h(k) - \left(\frac{4\pi a_h}{m_h}\right)^2 \sum_k (K_{hh} - K_{hh}^{vac}) G_h(k) + o(a_h p_{Fh})^3 \quad (19)$$

First term will get $\frac{4\pi a_h}{m_h} n_h$. It is just Hartree-Fock contribution. For the second term we can make an analytic continuation $i\omega_n \to \omega + io$ for bosonic propagator $K_{hh}$ and fermionic propagator $G_h$. As a result (having in mind that $\text{Im} K_{hh}^{vac} = 0$) we get for imaginary part of $\Sigma^{(2)}{}_{hh}$:

$$\operatorname{Im}\Sigma_{hh}^{(2)}(\varepsilon,\vec{p}) = \left(\frac{4\pi a_h}{m_h}\right)^2 \sum_k \operatorname{Im} K_{hh}(\varepsilon_k + \varepsilon, \vec{k}+\vec{p})[n_B(\varepsilon_k + \varepsilon) + n_F(\varepsilon_k)] =$$

$$= -\left(\frac{4\pi a_h}{m_h}\right)^2 \pi \int \frac{d^3\vec{k}}{(2\pi)^3} \int \frac{d^3\vec{p}'}{(2\pi)^3} \left[1 - n_h^F(\vec{p}+\vec{p}'+\vec{k}) - n_h^F(-\vec{p}')\right] [n_B(\varepsilon_k + \varepsilon) + n_F(\varepsilon_k)] \cdot \quad (20)$$

$$\cdot \delta\left[\varepsilon + \varepsilon_h(\vec{k}) - \varepsilon_h(\vec{p}+\vec{p}'+\vec{k}) - \varepsilon_h(-\vec{p}')\right]$$

and analogously for the real part of $\Sigma_{hh}^{(2)}$:

$$\operatorname{Re}\Sigma_{hh}^{(2)}(\varepsilon,\vec{p}) = \left(\frac{4\pi a_h}{m_h}\right)^2 \sum_k \left[\operatorname{Re} K_{hh}(\varepsilon_k + \varepsilon, \vec{k}+\vec{p}) - \operatorname{Re} K_{hh}^{vac}(\varepsilon_k + \varepsilon_p, \vec{k}+\vec{p})\right] [n_B(\varepsilon_k + \varepsilon) + n_F(\varepsilon_k)] \quad (21),$$

where for the real part of a Cooper loop in vacuum:

$$\operatorname{Re} K_{hh}^{vac}(\varepsilon_k + \varepsilon_p, \vec{k}+\vec{p}) = \int \frac{d^3\vec{p}'}{(2\pi)^3} P \frac{2m_h}{\vec{k}^2 + \vec{p}^2 - (\vec{p}'+\vec{k}+\vec{p})^2 - \vec{p}'^2} \quad (22)$$

is calculated in resonance for $\Omega = \varepsilon_k + \varepsilon_p$ (or for $\varepsilon = \varepsilon_p$), P-is principal value. In (20, 21) $n_B(\Omega) = \frac{1}{(e^{\Omega/T} - 1)}$ and $n_F(\Omega) = \frac{1}{(e^{\Omega/T} + 1)}$ - are bosonic and fermionic distribution functions and correspondingly:

$$n_B(\varepsilon_k + \varepsilon) + n_F(\varepsilon_k) = \frac{1}{2}\left[cth\frac{(\varepsilon_k + \varepsilon)}{2T} - th\frac{\varepsilon_k}{2T}\right] \quad (23)$$

The real part of a Cooper-loop in substance for heavy particles reads:

$$\operatorname{Re} K_{hh}(\varepsilon_k + \varepsilon, \vec{k}+\vec{p}) = \int \frac{d^3\vec{p}'}{(2\pi)^3} \frac{[1 - n_h^F(\vec{p}+\vec{p}'+\vec{k}) - n_h^F(-\vec{p}')]}{[\varepsilon + \varepsilon_h(\vec{k}) - \varepsilon_h(\vec{p}+\vec{p}'+\vec{k}) - \varepsilon_h(-\vec{p}')]}$$

The analytic continuation for $\Sigma^{(2)}{}_{hh}$ in a 2D case is similar to 3D case.

Note that for $\Omega/T \gg 1$ the bosonic distribution function $n_B(\Omega) \to 0$ and the fermionic distribution function $n_F(\Omega) \to \theta(\Omega)$ - step-function. As a result for $T \to 0$ $\operatorname{Im}\Sigma_{hh}$ and $\operatorname{Re}\Sigma_{hh}$ acquire the standard form [15,16,18]. In fact for low temperatures $T \ll W_h \ll W_L$ the most convenient way is to evaluate $\operatorname{Im}\Sigma^{(2)}{}_{hh}(\varepsilon)$ for $T \to 0$, thus get the standard Fermi-liquid result $\operatorname{Im}\Sigma^{(2)}{}_{hh}(\varepsilon) \sim \varepsilon^2$ and then make the temperature averaging with the corresponding fermionic distribution function $n_F(\varepsilon)$. Thus $\varepsilon \sim T$ for the lifetimes (or as we will show later for the scattering times) of the quasiparticles. The evaluation of $\Sigma_{hL}$, $\Sigma_{Lh}$ and $\Sigma_{LL}$ at low temperatures in first two orders in gas-parameter is similar to the evaluation of $\Sigma_{hh}$ both in 3D and 2D cases.

However for higher temperatures we should have in mind that $n_B(\Omega) \to T/\Omega$ for $T \gg \Omega$. The fermionic distribution function is "washed" out by temperature.

Accordingly $n_B(\Omega) = \frac{1}{2}\left(1 - \frac{\Omega}{2T}\right)$. These approximations are important when we evaluate $\text{Im}\Sigma$ for higher temperatures $T>W_h$ [20].

Note that in contrast with the model of slightly non-ideal Fermi-gas (see [15,16.18]) the Hubbard model does not contain an exchange-type diagram for $\Sigma_{hh}$ (see Fig. 4) since the T-matrix in this diagram corresponds to incoming and outgoing heavy particles with the same spin-projection $a^+_\sigma a^+_\sigma a_\sigma a_\sigma$ while the Hubbard model contains only the matrix elements $a^+_\uparrow a^+_\downarrow a_\downarrow a_\uparrow$.

Note also that when we expand the T-matrix till second order of gas-parameter we implicitly assume that the T-matrix itself does not have a simple pole-structure of a type of a bosonic propagator. This is a case for partially filled band $n_h d^D \ll 1$ and low energy sector where $0<\varepsilon<W_h\ll U_{hh}$. Effectively we neglect the lattice in this expansion.

However an account of the lattice produces two poles for the full (unexpanded) T-matrix of heavy particles in (15). First one is connected with the so-called antibound state predicted by Anderson[21] and corresponds to large positive energy

$\varepsilon \sim U_{hh} > 0$.           (24)

Physically it describes an antibound pair of two heavy particles with an energy $U_{hh}$ on the same lattice site. Thus it reflects the presence of the upper Hubbard band already at low densities $n_h d^D \ll 1$. However the intensity of the upper Hubbard band is small at low densities and for low energy sector.

Second pole in the full T-matrix found by Engelbrecht and Randeria [22] corresponds to negative energy and yields in 2D case:

$$\varepsilon \approx -2\varepsilon_{Fh} - \frac{2\varepsilon_{Fh}^2}{W_h} < 0 \quad (25)$$

It describes the bound state of the two holes below the bottom of the heavy band ($\varepsilon < -2\varepsilon_{Fh}$). Thus it has zero imaginary part and does not contribute to ImT. (This mode produces non-analytical corrections to $\text{Re}\Sigma_{hh} \sim |\varepsilon|^{5/2}$ in 2D). We can neglect both these two contributions for the self-energy when we will calculate the effective masses and lifetimes in the forthcoming sections. The more rigorous approach to the generalization of Galitskii results for the self-energy [15] on the case of finite temperatures (which is important for kinetic applications) will be a subject of separate publication.

## 5 Electron polaron effect

The Green-functions for heavy and light electrons for $T \to 0$ read:

$$G_h(\omega,\vec{q}) = \frac{1}{(\omega - \varepsilon_h(q) - \Sigma_h(\omega,\vec{q}))} \approx \frac{Z_h}{(\omega - \varepsilon_h^*(q) + io)}; \quad \text{and analogously} \quad (26)$$

$$G_L(\omega,\vec{q}) \approx \frac{Z_L}{(\omega - \varepsilon_L^*(q) + io)},$$

where $\varepsilon_h^*(q) = \dfrac{(q^2 - p_{Fh}^2)}{2m_h^*}$ and $\varepsilon_L^*(q) = \dfrac{(q^2 - p_{FL}^2)}{2m_L^*}$ (27)

are renormalized quasiparticle spectra;

$$Z_h^{-1} = \left(1 - \frac{\partial \operatorname{Re}\Sigma_h^{(2)}(\omega,\vec{q})}{\partial \omega}\bigg|_{\substack{\omega \to 0 \\ q \to p_{Fh}}}\right); Z_L^{-1} = \left(1 - \frac{\partial \operatorname{Re}\Sigma_L^{(2)}(\omega,\vec{q})}{\partial \omega}\bigg|_{\substack{\omega \to 0 \\ q \to p_{FL}}}\right) - \quad (28)$$

are Z-factors of heavy and light electrons. Substitution of the leading contribution from $\operatorname{Re}\Sigma_{hL}^{(2)}(\omega,\vec{q})$ (described by the formula similar to (21)) to $\operatorname{Re}\Sigma_h^{(2)}(\omega,\vec{q})$ in (28) yields:

$$\lim_{\substack{\omega \to 0 \\ q \to p_{Fh}}} \frac{\partial \operatorname{Re}\Sigma_{hL}^{(2)}(\omega,\vec{q})}{\partial \omega} \sim -\left(\frac{4\pi a_{hL}}{m_{hL}^*}\right)^2 \iint \frac{d^D\vec{p}}{(2\pi)^D}\frac{d^D\vec{p}\,'}{(2\pi)^D} \frac{[1 - n_L^F(\vec{p}\,'+\vec{p}) - n_h^F(-\vec{p}\,')]n_L^F(\vec{p}-\vec{q})}{[\varepsilon_L(\vec{p}-\vec{q}) - \varepsilon_L(\vec{p}\,'+\vec{p}) - \varepsilon_h(-\vec{p}\,')]^2}, \quad (29)$$

where $n_B(\Omega) \to 0$, $n_F(\Omega)$ is a step function for $\Omega/T \gg 1$; $a_{hL} \approx d$ is connected with the vacuum T-matrix $T_{hL}$; $m_{hL}^* \approx m_L$. Replacing in (29) $\dfrac{d^D\vec{p}}{(2\pi)^D}\dfrac{d^D\vec{p}\,'}{(2\pi)^D}$ by $N_L^2(0)d\varepsilon_L(\vec{p})d\varepsilon_L(\vec{p}\,')$ (where $N_L(0)$ is a density of states for light particles), and taking into account that $\varepsilon_L(\vec{p}-\vec{q}) < 0$ while $\varepsilon_L(\vec{p}\,'+\vec{p}) > 0$ we can easily check that for $m_h \gg m_L$ (or equivalently for $\varepsilon_{FL} \gg \varepsilon_{Fh}$) this expression contains a large logarithm (see [2]). Thus for Z-factor of the heavy particles in the leading approximation:

$$Z_h^{-1} \approx 1 + 2f_0^2 \ln\frac{m_h}{m_L}, \quad (30)$$

where $f_0 = 2p_{FL}d/\pi$ is the gas parameter in 3D and equivalently $f_0 = \dfrac{1}{2\ln(1/p_{FL}d)}$ in 2D. Note that the contribution to $Z_h$ from $\text{Re}\Sigma_{hh}^{(2)}$ does not contain a large logarithm. Correspondingly for effective mass of a heavy particle in (26) according to [16,18] we get:

$$\frac{m_h}{m_h^*} = Z_h \left( 1 + \frac{\partial \text{Re}\Sigma_{hL}^{(2)}(\varepsilon_h(\vec{q}),\vec{q})}{\partial \varepsilon_h(\vec{q})} \bigg|_{\varepsilon_h(q)\to 0} \right) \quad (31)$$

Thus, as usual, Z-factor contributes to the enhancement of a heavy mass:

$$\frac{m_h^*}{m_h} \sim Z_h^{-1} \sim \left( 1 + 2f_0^2 \ln\frac{m_h}{m_L} \right). \quad (32)$$

The analogous calculations for $Z_L$ with $\text{Re}\Sigma_{Lh}$ and $\text{Re}\Sigma_{LL}$ yields only $m_L^*/m_L \sim Z_L^{-1} \sim (1 + f_0^2)$. If the effective parameter $2f_0^2 \ln(m_h/m_L) > 1$ we are in the situation of strong electron polaron effect. In this region of parameters to get a correct polaron exponent diagrammatically we should at least sum up so-called maximally crossed diagrams for $\text{Re}\Sigma_{hL}$. The exponent evaluation could be fulfilled, however, in a different technique which is based on the non-adiabatic part of the many particle wave-function [2] which describes a heavy particle dressed in a cloud of electron – hole pairs of light particles. This yields:

$$\frac{m_h^*}{m_h} \sim Z_h^{-1} = \left( \frac{m_h}{m_L} \right)^{\frac{b}{(1-b)}}, \quad (33)$$

where $b = 2f_0^2$. For $b = \frac{1}{2}$ or equivalently for $f_0 = \frac{1}{2}$ (as for the coupling constant of the screened Coulomb interaction in the RPA-scheme) we are in the so-called unitary limit. In this limit according to [2] the polaron exponent is:

$$\frac{b}{(1-b)} = 1, \quad (34)$$

and thus:

$$\frac{m_h^*}{m_h} = \frac{m_h}{m_L} \quad (35)$$

or equivalently:

$$\frac{m_h^*}{m_L} = \left(\frac{m_h}{m_L}\right)^2. \tag{36}$$

Hence starting from the ratio between the bare masses $m_h/m_L \sim 10$ (obtained, for instance, in LDA-approximation) we finish in the unitary limit with $m_h^*/m_L \sim 100$ (due to many-body EPE), which is a typical ratio for uranium-based heavy-fermion (HF) systems.

**Other mechanisms of heavy mass enhancement**

Note that rigorously speaking (see (31)) the momentum dependence of $\text{Re}\Sigma_{hL}^{(2)}(\varepsilon_h(\vec{q}),\vec{q})$ is also very important for the evaluation of the effective mass. Very preliminary estimates of N.V. Prokof'ev and the author of a present paper [23] show that in zeroth approximation in $m_L/m_h$ in 3D case close to the Fermi surface (for $\varepsilon_h(q) = \frac{(q^2 - p_{Fh}^2)}{2m_h} \to 0$ and $q \to p_{Fh}$):

$$\text{Re}\Sigma_{hL}^{(2)}(\varepsilon_h(\vec{q}),\vec{q}) \approx 2\left(\frac{4\pi a_{hL}}{m_L}\right)^2 \int \frac{d^3\vec{p}}{(2\pi)^3} \Pi_{LL}(0,\vec{p}) n_h^F(\vec{p}-\vec{q}), \tag{37}$$

where

$$\Pi_{LL}(0,\vec{p}) = \int \frac{d^3\vec{p}'}{(2\pi)^3} \frac{[n_L^F(\varepsilon_{p'+p}) - n_L^F(\varepsilon_{p'})]}{\varepsilon_L(\vec{p}') - \varepsilon_L(\vec{p}'+\vec{p})} \tag{38}$$

is a static polarization operator for light particles. Having in mind that $|\vec{p}-\vec{q}| < p_{Fh}$; $q \approx p_{Fh}$ in (37) we can see that $\vec{p} \to 0$ and use the asymptotic form for $\Pi_{LL}(0,\vec{p})$ at small $p \ll p_{FL}$ (if the densities of heavy and light bands are not very different and $p_{FL} \sim p_{Fh}$):

$$\lim_{p>0} \Pi_{LL}(0,\vec{p}) = N_L(0)\left[1 - \frac{p^2}{12 p_{FL}^2}\right], \tag{39}$$

where $N_L(0) = m_L p_{FL}/2\pi^2$ is the density of states for light electrons in 3D. The substitution of $\lim_{p \to 0} \Pi_{LL}(0,\vec{p})$ from (39) to (37) yields:

$$\text{Re}\Sigma_{hL}^{(2)}(\varepsilon_h(\vec{q}),\vec{q}) \approx \text{Re}\Sigma_{hL}^{(2)}(0, p_{Fh}) - \frac{(q^2 - p_{Fh}^2)}{2m_h} \frac{f_0^2}{9} \frac{m_h n_h}{m_L n_L}, \qquad (40)$$

where $f_0 = 2dp_{FL}/\pi$ is a 3D gas parameter, $n_h = p_{Fh}^3/3\pi^2$, $n_L = p_{FL}^3/3\pi^2$ are the densities of heavy and light bands.

The first term in (40) describes $\text{Re}\Sigma_{hL}^{(2)}(\varepsilon_h(\vec{q}),\vec{q})$ on the Fermi surface (for $\varepsilon_h(q) = 0$ and $q = p_{Fh}$):

$$\text{Re}\Sigma_{hL}^{(2)}(0, p_{Fh}) \sim \frac{4f_0^2}{3} \frac{n_h}{n_L} \varepsilon_{FL} \left(1 - \frac{2p_{Fh}^2}{15 p_{FL}^2}\right) > 0 \text{ for } p_{Fh} \sim p_{FL}. \qquad (41)$$

It is renormalization of an effective chemical potential of the heavy band in the second order of the gas parameter due to the interaction of light and heavy particles.

Note that according to [15,16] the renormalized heavy-particle spectrum reads:

$$\varepsilon_h^*(q) = \left(\frac{q^2}{2m_h} - \mu_h^{eff}\right) + \frac{2\pi}{m_L} n_L(\mu) a_{hL} + \text{Re}\Sigma_{hL}^{(2)}(\varepsilon_h(\vec{q}),\vec{q}) = \frac{(q^2 - p_{Fh}^2)}{2m_h^*}, \qquad (42)$$

where the scattering length $a_{hL} \approx d$, an effective chemical potential $\mu_h^{eff} = \mu_h + W_h/2 + \varepsilon_0$ is counted from the bottom of a heavy band, and the Hartree-Fock term $(2\pi/m_L)n_L(\mu)a_{hL}$ represents the first-order in gas parameter contribution to the self-energy $\text{Re}\Sigma_{hL}^{(1)}$. Thus collecting the terms proportional to $\varepsilon_h(q) = \frac{(q^2 - p_{Fh}^2)}{2m_h}$ we get from (42):

$$\frac{(q^2 - p_{Fh}^2)}{2m_h^*} = \varepsilon_h(q)\left[1 - \frac{f_0^2}{9} \frac{m_h n_h}{m_L n_L}\right] \qquad (43)$$

Correspondingly the effective mass of a heavy particle is given by:

$$\frac{m_h}{m_h^*} = \left(1 + \frac{\partial \text{Re}\Sigma_{hL}^{(2)}(\varepsilon_h(q),\vec{q})}{\partial \varepsilon_h(q)}\bigg|_{\varepsilon_h \to 0}\right) = \left(1 - \frac{f_0^2}{9} \frac{m_h n_h}{m_L n_L}\right). \qquad (44)$$

As a result we get much more dramatic enhancement of $m_h^*$ than EPE which yields only $m_h/m_h^* \approx (1 - 2f_0^2 \ln(m_h/m_L))$ via Z-factor of a heavy particle. Note that the contribution to $m_h^*/m_h$ from $\text{Re}\Sigma_{hh}^{(2)}(\varepsilon_h(q),\vec{q})$ connected with "heavy-heavy" interaction is small in comparison with the contribution to $m_h^*$ from Re$\Sigma_{hL}^{(2)}$ (which is connected with "heavy-light" interaction) due to the smallness of the ratio between the bare masses: $m_L/m_h \ll 1$. Now we can collect the terms which do not depend upon $\varepsilon_h(q)$ in (42). Thus we get for the effective chemical potential of heavy electrons:

$$\mu_h^{eff} = \frac{p_{Fh}^2}{2m_h} + \frac{2\pi}{m_L} n_L(\mu) a_{hL} + \text{Re}\Sigma_{hL}^{(2)}(0, p_{Fh}) \qquad (45)$$

Note that the contributions to $\mu_h^{eff}$ from the Hartree-Fock term $(2\pi/m_h)n_h(\mu)a_{hh}$ of heavy electrons and from Re$\Sigma_{hh}^{(2)}(0,p_{Fh})$ (which is connected with "heavy-heavy" interactions) are small in comparison with "heavy-light" contributions due to the smallness of the ratio between the bare masses: $m_L/m_h \ll 1$.

In 2D the static polarization operator reads: $\Pi_{LL}(0, \vec{p}) = \frac{m_L}{2\pi}\left[1 - \text{Re}\sqrt{1 - \frac{4p_{FL}^2}{p^2}}\right]$ and thus for $p < 2p_{FL}$: $\Pi_{LL}(0, \vec{p}) = \frac{m_L}{2\pi}$ - does not contain any dependence upon $p^2$ in contrast with 3D-case. Thus in 2D EPE is a dominant mechanism of the heavy mass enhancement.

More accurate evaluation of momentum dependence of $\text{Re}\Sigma_{hL}^{(2)}(\varepsilon_h(q),\vec{q})$ for the larger densities in the bands together with the summation of the higher order contributions to Re$\Sigma_{hL}$ will be a subject of a separate investigation.

Note that for the light particles momentum dependences of Re$\Sigma_{Lh}^{(2)}$ and Re$\Sigma_{LL}^{(2)}$ yield only $m_L^*/m_L \sim 1 + f_0^2$ and thus the light mass is not strongly enhanced both in 3D and 2D cases.

**The tendency towards phase-separation**

Note also that for larger densities of the heavy band $n_h \sim n_C \leq 1$ (and large difference in densities between the bands: $n_L \ll n_h$, so $n_{tot} = n_h + n_L \leq 1$) another mechanisms of heavy mass enhancement become more effective. Namely for these densities and large mismatch between $n_h$ and $n_L$ we could have a tendency towards phase-separation in a two-band model [11].

Note that if we analyze the effective chemical potential of the heavy band (45) in the limit of the large density mismatch $n_h \gg n_L$ in 3D and evaluate the partial compressibility (sound velocity squared of heavy particles) $\kappa_{hh}^{-1} \sim c_h^2 = \frac{n_h}{m_h}\left(\frac{\partial \mu_h}{\partial n_h}\right)$ we already see the tendency towards phase-separation (towards negative compressibility) in the strong coupling limit and low densities for $f_0^2 \frac{m_h p_{Fh}}{m_L p_{FL}} \geq 1$ in qualitative agreement with the results of [11]. The more careful analysis of all the partial compressibilities in the system at larger $f_0$ and large mismatch between the densities will be a subject of the separate publication.

In the end of this Section we would like to emphasize that the physics of EPE and evaluation of $Z_h$ in [2] is to some extent connected with the well-known results of P. Nozieres et al., [24] on infrared divergences in the description of the Brownian motion of a heavy particle in a Fermi liquid and on the infrared divergences for the problem of X-ray photoemission from the deep electron levels, as well as with the famous results of P.W. Anderson [25] on the orthogonality catastrophe for the 1D chain of $N$ electrons under the addition of one impurity to the system.

Finally I would like to mention here a competing mechanism of P. Fulde et al., [26] worked out firstly for the explanation of the effective mass in praseodymium (Pr) and in some uranium-based molecules like $U(C_8H_8)_2$. Later on P. Fulde et al., generalized this mechanism on some other uranium-based HF-compounds with localized and delocalized orbitals. This mechanism has a quantum-chemical nature and is based on the scattering of conductive electrons on localized orbitals as if on the two-level systems. The mass enhancement here is governed by non-diagonal matrix elements of Coulomb interaction which are not contained in the simple version of a two-band model (1). In this context we would like to mention also [27] where the authors considered the mass enhancement of conductivity electrons due to their scattering on local $f$-levels splitted by crystalline field.

Note that dHvA-experiments [28] together with ARPES-experiments [29] and thermodynamic measurements [30] are the main instruments to reconstruct the Fermi surface for HF-compounds and to determine the effective mass (thus verifying the predictions of different theories on the mass enhancement in uranium-based HF-compounds).

# 6 The temperature dependence of the resistivity

**Imaginary parts of the self-energies in the homogeneous state**

For $T \to 0$ all the imaginary parts of the self-energies in the homogeneous state for equal densities of heavy and light electrons behave in a standard FL-manner. For $\varepsilon_q > 0$ they read:

$$\mathrm{Im}\Sigma_{hh}^{(2)}(\varepsilon_h(\vec{q}),\vec{q}) = f_0^2 \frac{\varepsilon_h^2(\vec{q})}{\varepsilon_{Fh}}; \quad \mathrm{Im}\Sigma_{LL}^{(2)}(\varepsilon_L(\vec{q}),\vec{q}) = f_0^2 \frac{\varepsilon_L^2(\vec{q})}{\varepsilon_{FL}}. \tag{46}$$

Accordingly for $\Sigma_{hL}$ and $\Sigma_{Lh}$ we get:

$$\mathrm{Im}\Sigma_{hL}^{(2)}(\varepsilon_h(\vec{q}),\vec{q}) = f_0^2 \frac{\varepsilon_h^2(\vec{q})}{\varepsilon_{Fh}}; \quad \mathrm{Im}\Sigma_{Lh}^{(2)}(\varepsilon_L(\vec{q}),\vec{q}) = f_0^2 \frac{\varepsilon_L^2(\vec{q})}{\varepsilon_{Fh}} \frac{m_h}{m_L}. \tag{47}$$

Note that $n_B(\Omega) \to 0$ and $n_F(\Omega) \to \theta(\Omega)$ for $\Omega/T \gg 1$ in the general expression for $\mathrm{Im}\Sigma^{(2)}$ obtained in Sec. 4.

**The scattering times and Drude conductivities**

For the inverse scattering times (more rigorously for the lifetimes) of the heavy and light particles for $\varepsilon \sim T$ we get from $\mathrm{Im}\Sigma^{(2)}$ in (46), (47):

$$1/\tau_h = (1/\tau_{hh} + 1/\tau_{hL}) = f_0^2 (T^2/\varepsilon_{Fh}). \tag{48}$$

Analogously for light particles:

$$1/\tau_L = (1/\tau_{LL} + 1/\tau_{Lh}) \approx 1/\tau_{Lh} \approx f_0^2 \frac{T^2}{\varepsilon_{Fh}} \frac{m_h}{m_L} > 1/\tau_h. \tag{49}$$

Now we can calculate the Drude conductivities according to the standard formulas $\sigma = ne^2\tau/m$. For light electrons:

$$\sigma_L = \frac{n_L e^2 \tau_L}{m_L} = \frac{n_L e^2}{f_0^2 T^2} \frac{\varepsilon_{Fh} m_L}{m_h m_L} \sim \frac{n_L e^2}{f_0^2 p_{Fh}^2} \left(\frac{\varepsilon_{Fh}}{T}\right)^2. \tag{50}$$

Introducing the minimal Mott – Regel conductivities:

$$\sigma_{\min} = (e^2/\hbar) p_F \text{ in 3D and } \sigma_{\min} = e^2/\hbar \text{ in 2D}, \tag{51}$$

and working in the units where $\hbar = 1$ we get for equal densities of heavy and light bands $n_L = n_h$:

$$\sigma_L = \frac{\sigma_{min}}{f_0^2}\left(\frac{\varepsilon_{Fh}}{T}\right)^2. \qquad (52)$$

Analogously for $\sigma_h$:

$$\sigma_h = \frac{n_h e^2 \tau_h}{m_h} = \frac{n_h e^2}{m_h}\frac{\varepsilon_{Fh}}{f_0^2 T^2} \sim \frac{n_h e^2}{p_{Fh}^2}\left(\frac{\varepsilon_{Fh}}{T}\right)^2 \frac{1}{f_0^2}. \qquad (53)$$

Thus the scattering times for heavy and light particles $1/\tau_h$ and $1/\tau_L$ differ, but the conductivities $\sigma \sim \tau/m$ have the same order of magnitude [2]:

$$\sigma_h \sim \frac{\sigma_{min}}{f_0^2}\left(\frac{\varepsilon_{Fh}}{T}\right)^2 \sim \sigma_L. \qquad (54)$$

The total conductivity reads:

$$\sigma = \sigma_h + \sigma_L \sim \frac{\sigma_{min}}{f_0^2}\left(\frac{\varepsilon_{Fh}}{T}\right)^2 \qquad (55)$$

and hence the resistivity:

$$R = \frac{1}{\sigma} = \frac{f_0^2}{\sigma_{min}}\left(\frac{T}{\varepsilon_{Fh}}\right)^2 \qquad (56)$$

behaves in a Fermi liquid manner $R(T) \sim T^2$ at low temperatures.

**The difference between lifetimes and transport times**

Rigorously speaking we calculate lifetimes and not transport times. However an exact solution of coupled kinetic equations [31] for heavy and light electrons with an account of umklapp processes for not too small densities of the bands shows that for $m_h \gg m_L$ and for $p_{Fh} \sim p_{FL} \sim p_F \leq 1/d$ for all the times including $\tau_{Lh}$, $\tau_{hL}$ we get [20]:

$$\tau_{transp} = \tau_{life\text{-}time} \qquad (57)$$

Note that umklapp processes for the interaction of heavy and light electrons imply:

$$\vec{p}_{1h} + \vec{p}_{2L} = \vec{p}_{3h} + \vec{p}_{4L} + \vec{K}, \quad (58)$$

where $K \sim \pi/d$ – is the wave-vector of the reciprocal lattice. For $p_{Fh} \sim p_{FL}$ it means that densities in light and heavy bands cannot be very small (otherwise resistivity will be exponentially small). Hence within the accuracy of our estimates:

$$R \sim \frac{f_0^2}{\sigma_{min}} \left(\frac{T}{W_h}\right)^2 \quad (59)$$

and in all the estimates for inverse scattering times and conductivities we can replace $\varepsilon_{Fh}$ on $W_h$ and $\varepsilon_{FL}$ on $W_L$. Moreover for $m_h^*/m_h \gg 1$ we can replace $m_h$ on $m_h^*$ (or equivalently $W_h$ on $W_h^*$) and thus the final result for the resistivity reads:

$$R \sim \frac{f_0^2}{\sigma_{min}} \left(\frac{T}{W_h^*}\right)^2. \quad (60)$$

**The chemical potential at higher temperatures $T > W_h^*$**

If $T > W_h^*$ the heavy band is totally destroyed (more precisely it is destroyed for $f_0^2 T = W_h^*$ as we will see soon). To be accurate let us first calculate the effective chemical potential $\mu_h^{eff} = \mu + \frac{W_h}{2} + \varepsilon_0$ in (3) in this situation.

Generally speaking $n_h + n_L = n_{tot} = const$ – only a total density is conserved. In our case however for large difference between the bare masses $m_h \gg m_L$, each density of the band is conserved practically independently $n_h \approx const$, $n_L \approx const$. For heavy particles all the states in the band will be uniformly occupied at these temperatures. For $T > W_h$ (assuming $m_h^*/m_h \sim 1$) an effective chemical potential of the heavy particles reads:

$$\mu_h^{eff} = \mu + \frac{W_h}{2} + \varepsilon_0 \sim -T \ln\left(\frac{1}{n_h d^D}\right). \quad (61)$$

Thus we have Boltzman behavior for $\mu_h^{eff}$. The Fermi – Dirac distribution function for heavy particles:

$$n_h(\varepsilon) = \frac{1}{(e^{p^2/2m_h - \mu_h^{eff}}{T} + 1)} \approx \frac{1}{\left[\left(1 + \frac{p^2}{2m_h T}\right)e^{-\mu_h^{eff}/T} + 1\right]} \approx \frac{e^{\mu_h^{eff}/T}}{\left(1 + \frac{p^2}{2m_h T}\right)} \approx e^{\mu_h^{eff}/T} = const \quad (62)$$

For light particles for the temperatures $W_h \ll T \ll W_L$ since $m_h \gg m_L$ an effective chemical potential will be approximately in the same place as for $T = 0$. Indeed for $\mu_{eff}^L = \mu + \frac{W_L}{2}$ we get:

$$n_L(\varepsilon) = \frac{1}{(e^{p^2/2m_L - \mu_h^{eff}}{T} + 1)} \approx \frac{1}{(e^{(p^2 - p_{FL}^2)/2m_L T} + 1)} \approx \theta\left(\frac{p^2}{2m_L} - \varepsilon_{FL}\right) \quad \text{for } T \ll \varepsilon_{FL} \quad (63)$$

and hence for the effective chemical potential of light particles:

$$\mu_{eff}^L \approx \varepsilon_{FL}. \quad (64)$$

**Evaluation of the scattering times at higher temperatures $W_h^* < T < W_L$**

For light particles the scattering time $1/\tau_{LL} = f_0^2 T^2/W_L$ – does not change. However:

$$\frac{1}{\tau_{Lh}} = f_0^2 W_h \frac{m_h}{m_L} - \quad (65)$$

- almost elastic scattering of light electrons on the heavy ones as if on immobile (static) impurities in zeroth order in $W_h/W_L$. Note that $W_h m_h = W_h^* m_h^*$.

For heavy electrons we should take into account that bosonic contribution $n_B(\Omega) \approx T/\Omega$ and fermionic contribution $n_F(\Omega) \approx \frac{1}{2}(1 - \Omega/2T)$ for $\Omega/T \ll 1$ in $\text{Im}\Sigma^{(2)}$ and thus in scattering times. This yields:

$$\frac{1}{\tau_{hh}} = f_0^2 W_h \quad (66)$$

- scattering of heavy electrons on each other in the situation when they uniformly occupy the heavy band and can transfer to each other only the energy $\sim W_h$ [2]. For $m_h^* \gg m_h$ we can replace $W_h$ on $W_h^*$ in (66). In the same time:

$$\frac{1}{\tau_{hL}} = f_0^2 T \quad (67)$$

marginal Fermi liquid behavior for diffusive motion of heavy electrons in the surrounding of light electrons.

**Resistivity for $T > W_h^*$ in a 3D case**

Hence for scattering times of heavy and light particles for $T > W_h^*$:

$$1/\tau_L = 1/\tau_{LL} + 1/\tau_{Lh} \approx 1/\tau_{Lh} = f_0^2 W_h \frac{m_h}{m_L} \qquad (68)$$

(for $T < W_L$: $T^2/W_L < W_h\, m_h/m_L$). Note that $f_0^2 W_h \frac{m_h}{m_L} = f_0^2 W_h^* \frac{m_h^*}{m_L} \sim f_0^2 W_L$ in (68).

In the same time:

$$1/\tau_h = 1/\tau_{hh} + 1/\tau_{hL} \approx 1/\tau_{hL} = f_0^2 T \qquad (69)$$

So the heavy component is marginal while the light component is not.
For conductivity of the light band:

$$\sigma_L = \frac{n_L e^2 \tau_L}{m_L} \approx \frac{n_L e^2 \tau_{Lh}}{m_L} = \frac{\sigma_{\min}}{f_0^2} \qquad (70)$$

For heavy band the Drude formula should be modified since for $T > W_h^*$: $\frac{\partial n_h}{\partial T} \sim \frac{W_h^*}{T}$. Then we immediately obtain:

$$\sigma_h = \frac{\sigma_{\min}}{f_0^2}\left(\frac{W_h^*}{T}\right)^2 \qquad (71)$$

As a result for the resistivity:

$$R = \frac{1}{(\sigma_h + \sigma_L)} = \frac{f_0^2}{\sigma_{\min}} \frac{(T/W_h^*)^2}{1 + (T/W_h^*)^2} = \frac{f_0^2}{\sigma_{\min}} \frac{1}{[1 + (W_h^*/T)^2]} \qquad (72)$$

For $T > W_h^*$ the resistivity $R \approx f_0^2/\sigma_{\min}$ – goes to saturation. So we obtain residual resistivity at high temperatures due to conductivity of a light band. It is a very nontrivial result.

**Discussion of the obtained results for resistivity at higher temperatures**

When $W_h^* < 1/\tau_h$ or equivalently $f_0^2 T > W_h^*$ the coherent motion in the heavy band is totally destroyed. The heavy particles begin to move diffusively in the surrounding of light particles. In this regime, rigorously speaking, the diagrammatic technique can be used only for light particles and not for the heavy ones.

However exact solution for density matrix equation obtained in [2] shows that $1/\tau_{hL}$ is qualitatively the same for $f_0^2 T > W_h^*$ as in our estimates, the inverse scattering time $1/\tau_{Lh}$ is also qualitatively the same due to its physical meaning (scattering of light electrons on heavy ones as if on immobile impurities). That is why $\sigma_h$, $\sigma_L$ and hence $R(T)$ behave smoothly for $f_0^2 T \geq W_h^*$.

**An idea of a hidden heavy band for HTSC**

The resistivity characteristics $R(T)$ in 3D acquires a form (see Fig. 5) which is frequently obtained in uranium-based HF (for example in $UNi_2Al_3$). Note that $R(T)$ mimics linear behavior in a crossover region of intermediate temperatures $T \sim W_h^*$ between $T^2$ and $const$ (where it goes on saturation for $T \gg W_h^*$). The same holds for magnetoresistance in the well – known experiments of P.L. Kapitza:

$$\frac{R(H)-R(0)}{R(H)} \sim \frac{(\Omega_C \tau)^2}{1+(\Omega_C \tau)^2} \sim \begin{cases} (\Omega_C \tau)^2 \text{ for } \Omega_C \tau < 1 \\ const \text{ for } \Omega_C \tau > 1 \end{cases}, \quad (73)$$

where $\Omega_C$ – is cyclotron frequency.

In the crossover region $\Omega_C \tau \sim 1$ magnetoresistance mimics linear in $\Omega_C$ behavior. Thus we obtain that for $T > W_h^*$ heavy electrons are marginal but light electrons are not. The natural question arises: whether it is possible to make light electrons also marginal and as a result to get the resistivity characteristics of the type: $R(T) \sim T$ – marginal for $T > W_h^*$ while $R(T) \sim T^2$ for $T < W_h^*$. Such resistivity characteristics could serve as an alternative scenario for the explanation of the normal properties in optimally doped or slightly overdoped HTSC – materials if we assume an existence of a hidden heavy band with a bandwidth smaller than a superconductive critical temperature $T_C$: $W_h^* < T_C$ (see Fig. 6). Then to get $R(T) \sim T^2$ – FL-behavior at low temperatures we should suppress SC by a large magnetic field $H$ till low critical temperatures $T_C(H) < W_h^*$.

# 7 Weak-localization corrections in a 2D case

The tendency towards marginalization of light component manifests itself in 2D case. We know that in 2D there are logarithmic corrections [3] due to weak localization effects to the classical Drude formula for conductivity. But according to our ideology heavy particles play the role of impurities for scattering of light particles on them. That is why the correct expression of conductivity of the light band $\sigma_L$ in the absence of spin-orbital coupling reads:

$$\sigma_L^{loc} = \frac{\sigma_{min}}{f_0^2}\left[1 - f_0^2 \ln\frac{\tau_\varphi}{\tau}\right], \quad (74)$$

where, according to weak-localization theory in 2D, $\tau$ is elastic time, while $\tau_\varphi$ is inelastic (decoherence) time. In our case:

$$\tau = \tau_{ei} = \tau_{Lh}, \text{ while } \tau_\varphi = \tau_{ee} = \tau_{LL}, \text{ and } \tau_{LL} \gg \tau_{Lh}, \quad (75)$$

where $\tau_{ei}$ and $\tau_{ee}$ – are the times connected with the scattering of electrons on impurities and other electrons respectively. Thus between two scatterings of a light particle on a light one it scatters a lot of time on heavy particles (see Fig. 7). As a result a motion of the light particles becomes much more slow (also of the diffusive type) and two characteristic lengths appear in the theory:

$$l = v_{FL}\tau_{Lh} \quad (76)$$

elastic length and:

$$L_\varphi = \sqrt{D_L \tau_\varphi} \quad (77)$$

- diffusive length, where $D_L$ is a diffusion coefficient for light electrons and $v_{FL}$ is Fermi velocity for light electrons.

That is why according to Altshuler-Aronov [3] in a more rigorous theory we should replace the inverse scattering time:

$$\frac{1}{\tau_{LL}(\varepsilon)} \sim \int_0^\varepsilon d\omega \int_0^\omega d\varepsilon' \int_0^\infty \frac{a_{LL}^2}{m_L^2} \frac{q\,dq}{(v_{FL}q)^2} = f_0^2 \frac{T^2}{W_L} \quad (78)$$

on:

$$\frac{1}{\tilde{\tau}_{LL}(\varepsilon)} \sim \int_0^\varepsilon d\omega \int_0^\omega d\varepsilon' \int_0^\infty \frac{a_{LL}^2}{m_L^2} \frac{q\,dq}{(i\varepsilon' + D_L q^2)^2} \quad (79)$$

where the scattering length $a_{LL} \sim d$. In fact we substitute $v_{FL}q$ by the "cooperon" pole $(i\varepsilon + D_L q^2)$ in Altshuler – Aronov terminology. Thus in the evaluation of $\tilde{\tau}_{LL}$ characteristic wave-vectors $q \sim \sqrt{\varepsilon/D_L}$, where $\varepsilon$ is an energy variable. Altshuler-Aronov effect in 2D yields:

$$\frac{1}{\tilde{\tau}_{LL}(\varepsilon)} = f_0^2 \frac{\varepsilon}{N_L(0)D_L}, \qquad (80)$$

where $N_L(0) = m_L/2\pi$ is a 2D density of states for light electrons. For diffusion coefficient we can use an estimate:

$$D_L = v_{FL}^2 \tau_{Lh} \qquad (81)$$

and hence having in mind that according to (69) the inverse scattering time $\frac{1}{\tau_{Lh}(\varepsilon)} = f_0^2 W_h \frac{m_h}{m_L} \approx f_0^2 W_L$ we get:

$$\frac{1}{\tilde{\tau}_{LL}(\varepsilon)} \sim \frac{f_0^2 f_0^2 W_L}{\frac{m_L}{\pi} v_{FL}^2} \varepsilon \sim f_0^4 \varepsilon \qquad (82)$$

Thus $1/\tilde{\tau}_{LL}$ also becomes marginal for $\varepsilon \sim T$. For logarithmic corrections to conductivity we have:

$$\frac{\tau_\varphi}{\tau} = \frac{\tilde{\tau}_{LL}}{\tau_{Lh}} = \frac{W_L}{f_0^2 T} \gg 1 \qquad (83)$$

and hence:

$$\sigma_L^{loc} = \frac{\sigma_{min}}{f_0^2}\left[1 - f_0^2 \ln\frac{W_L}{f_0^2 T}\right]. \qquad (84)$$

For $f_0^2 T \sim W_h$: $\ln\frac{W_L}{f_0^2 T} \sim \ln\frac{W_L}{W_h}$ and

$$Z_h = \frac{\sigma_L^{loc}}{\sigma_L} = 1 - f_0^2 \ln\frac{W_L}{W_h}. \qquad (85)$$

So for $f_0^2 T \sim W_h$ an enhancement of a heavy particle Z-factor due to EPE and localization of light particles due to Altshuler-Aronov corrections are governed by the same parameter $f_0^2 \ln(m_h/m_L)$ in 2D.

**Justification of the expression for localization – corrections in 2D**

In principle impurities (heavy particles) are mobile and have some recoil energy. That is why the formula $\dfrac{\sigma_L^{loc}}{\sigma_L} = 1 - f_0^2 \ln \dfrac{W_L}{f_0^2 T}$ should be justified (at least temperature exponent under logarithm: $T$ or $T^\alpha$). For the justification we need to estimate the loss of energy by one light particle before it collides with another light particle. Number of collisions with heavy particles in between the scattering of light particle on a light one is $L_\varphi/l$. Maximal loss of energy in one collision is $W_h^*$. Total loss is $W_h^* \dfrac{L_\varphi}{l} = W_h^* \sqrt{\dfrac{W_L}{T}}$. The energy of light particle itself is $T$. It means that for $W_h^* \sqrt{\dfrac{W_L}{T}} < T$ or equivalently for

$$T > W_h^* \left( \dfrac{W_L}{f_0^2 W_h^*} \right)^{1/3}. \qquad (86)$$

the loss of energy is small and heavy particles can be regarded as immobile impurities. Thus exponent $\alpha$ under logarithm is 1.

**The resistivity in a 2D case**

Qualitatively resistivity behaves in 2D in the following manner:

$$R = \dfrac{f_0^2}{\sigma_{\min}} \dfrac{1}{\left[ \left( \dfrac{W_h^*}{T} \right)^2 + \left( 1 - f_0^2 \ln \dfrac{W_L}{f_0^2 T} \right) \right]}. \qquad (87)$$

It has a maximum for $T_{\max} \sim W_h^*/f_0$ and localization tail at higher temperatures (see Fig. 8). It will be very interesting to find magnetoresistance in the 2D or layered case in a two-band model with one narrow band for a strong quantizing magnetic field $H$ oriented perpendicular to the layers. We can expect here a strong manifestation of famous Aharonov – Bohm effect [32].

# 8 Superconductivity in the two-band model with one narrow band

In the end of this paper let us mention briefly that the leading mechanism of SC at low electron density corresponds to p-wave pairing and is governed, especially in 2D, by the pairing of heavy electrons via polarization of light ones (see Fig.9 and [12,13]) in the framework of enhanced Kohn-Luttinger mechanism[4]. The corresponding $T_c$ nonmonotonically depends upon the relative doping of the bands $n_h/n_L$ and has a broad and pronounced maximum for $n_h/n_L = 4$ in 2D, where it could reach the experimentally feasible values realistic for layered ruthenates $Sr_2RuO_4$ [38] and uranium-based heavy-fermion compounds like $U_{1-x}Th_xBe_{13}$ [37] as well as for layered dichalcogenides $CuS_2$, $CuSe_2$ and semimetallic superlattices InAs-GaSb, PbTe-SnTe with geometrically separated bands belonging to different layers [36]

In the situation of weak EPE for $f_0^2 \ln \frac{m_h}{m_L} < 1$, $Z_h \sim \frac{m_h}{m_h^*} \sim 1$, $\varepsilon_{Fh}^* \sim \varepsilon_{Fh}$ and accordingly to [12,13,20] the maximal $T_{C1}$ reads:

$$T_{C1} \sim \varepsilon_{Fh} \exp\left\{-\frac{1}{2f_0^2 \frac{m_h}{m_L}}\right\} \qquad (88)$$

Thus effective gas-parameter which governs $T_{C1}$ in case of weak EPE is $f_0 \left(\frac{m_h}{m_L}\right)^{1/2}$. In the same time in the unitarian limit for $f_0 \to 1/2$ and $m_h^*/m_L \sim (m_h/m_L)^2$ the estimates show that:

$$T_{C1} \sim \varepsilon_{Fh}^* \exp\left\{-\frac{1}{2f_0^2}\right\} \sim \varepsilon_{Fh}^* \exp\{-2\} \quad (90)$$

Thus for $\varepsilon_{Fh}^* \sim 50K$ $T_{C1}$ could reach 5K which is quite nice.

When we increase the density of a heavy band and go closer to half-filling ($n_h \to 1$) the d-wave SC-paring (as in UPt$_3$) becomes more beneficial in the framework of the spin-fluctuation theory in the heavy-band [39,40]. The more exotic mechanisms of SC in heavy-fermion compounds including odd-frequency pairing [5,41] are also possible.

Note that in 2D case, where only EPE-effect is present for the mass-enhancement of heavy electrons, the restrictions on a homogeneous case are more mild then in 3D.

## 9 Discussion and Conclusions

We analyzed characteristic features of the two-band Hubbard model with one narrow band taking into account electron – electron scattering in a clean case (no impurities) and low electron densities. We considered electron polaron effect and other mechanisms of a heavy mass enhancement related to momentum dependence of the self-energies.

In the 3D-case the dominant mechanism of a heavy mass enhancement is related to momentum-dependence of a real part of a "heavy – light" self-energy and leads to linear in the mass-ratio renormalization of a heavy mass. In the 2D-case the dominant mechanism of a heavy mass enhancement is EPE which leads to logarithmic renormalization of the heavy particle Z-factor. In the unitary limit if we start with $m_h/m_L \sim 10$ for the bare-mass ratio in the LDA-scheme we can finish with $m_h^*/m_L \sim 100$ due to many-body effects which is quite natural for uranium-based HF systems.

The important role of the interband ("heavy – light") Hubbard repulsion $U_{hL}$ for the formation of a heavy mass $m^* \sim 100\, m_e$ in a two-band Hubbard model was also emphasized in [33] for LiV$_2$O$_4$ HF compound.

For large density mismatch $n_h \gg n_L$ we can see the tendency towards negative compressibility in a heavy band in a strong coupling limit $f_0^2 \dfrac{m_h p_{Fh}}{m_L p_{FL}} \geq 1$ already at low densities, which can lead to the redistribution of charge between the bands and possibly to nanoscale phase-separation in qualitative similarity with the results of [11]. The tendency towards phase-separation at low electron fillings also manifests itself for the asymmetric Hubbard model (which possesses Hubbard repulsion between heavy and light electrons) in the limit of strong asymmetry: $t_h \ll t_L$ [34] between heavy and light bandwidths.

For equal densities of heavy and light bands the resistivity in a homogeneous state behaves in a Fermi – liquid fashion: $R(T) \sim T^2$ at low temperatures $T < W_h^*$ both in 3D and in 2D cases (where $W_h^*$ is an effective bandwidth of heavy particles).

For higher temperatures $T > W_h^*$ when a coherent motion of particles in a heavy band is totally destroyed, the heavy particles move diffusively in the surrounding of light particles while the light particles scatter on the heavy ones as if on immobile (static) impurities. The resistivity goes on saturation in 3D-case which is typical for some uranium-based HF-compounds including $UNi_2Al_3$.

In 2D due to weak-localization corrections of Altshuler-Aronov type the resistivity at higher temperatures has a maximum and then a localization tail. Such behavior could be also relevant for some other mixed-valence systems possibly including layered manganites. The similar behavior with metal-like low temperature dependence of resistivity for $T < 130$ K and insulator-like high-temperature dependence was also observed in layered intermetallic alloys $Gd_5Ge_4$ where the authors [35] assume an existence of strongly-correlated narrow band at low temperatures.

We discuss briefly the SC-instabilities which arise in this model at low electron densities. The leading instability of the enhanced Kohn – Luttinger type corresponds to p-wave pairing of heavy electrons via polarization of light electrons. In quasi-2D case $T_C$ can reach experimentally realistic values already at low densities for layered dichalcogenides $CuS_2$, $CuSe_2$ and semimetallic superlattices InAs-GaSb, PbTe-SnTe with geometrically separated bands belonging to neighboring layers [36]. Note that p-wave SC is widely discussed in 3D heavy-fermion systems like $U_{1-x}Th_xBe_{13}$ [37] and in layered ruthenates $Sr_2RuO_4$ with several pockets (bands) for conducting electrons [38]. Note also that when we increase the density of a heavy-band and go closer to half-filling ($n_h \to 1$) the d-wave superconductive pairing (as in $UPt_3$) becomes more beneficial in the framework of the spin-fluctuation theory in the heavy band [39,40]. Different mechanisms of SC in HF-compounds including odd-frequency pairing are discussed in [41] by P. Coleman et al.

Note also that if we study the orbitally degenerate two-band Hubbard model then Hubbard parameters read $U = U_{hh} = U_{LL} = U_{hL} + 2J_H$ (where $J_H$ is Hund's coupling) [42]. Close to half-filling this model becomes equivalent to the *t-J* orbital model [43] and contains for $J < t$ and at optimal doping the SC d-wave pairing [44] governed by superexchange interaction between the different orbitals of AFM-type $J > 0$. Note that for not very different values of $t_h$ and $t_L$ the typical value of $J \sim t^2/U \sim 300$ K. The orbital *t-J* model also reveals a tendency towards nanoscale phase-separation at low doping [45] with the creation of orbital ferrons inside insulating AFM orbital matrix. An orbital type of phase-separation was possibly observed in $URu_2Si_2$ [46].

**Acknowledgements**   We are grateful to A.S. Alexandrov, A.F. Andreev, A.F. Barabanov, M.A. Baranov, Yu. Bychkov, A.V. Chubukov, D.V. Efremov, P. Fulde, A.S. Hewson, Yu. Kagan, K.A. Kikoin, K.I. Kugel, F.V. Kusmartsev, Yu. E. Lozovik, P. Nozieres, N.V. Prokof'ev, A.L. Rakhmanov, T.M. Rice, A.O. Sboychakov, P. Thalmeer, C.M. Varma, D. Vollhardt, P. Woelfle, A. Yaresko for a lot of simulating discussions on this subject and acknowledge financial support of the RFBR grants # 08-02-00224, 08-02-00212. M.Yu.K. is also grateful to Leverhulme trust for granting the visit to Loughborough University, where this work was completed.

**References**

1. C.M. Varma, P.B. Littlewood, S. Schmitt-Rink, E. Abrahams, A.E. Ruchenstein, Phys. Rev. Lett., 63, 1996 (1989)

2. Yu. Kagan, N.V. Prokof'ev, Sov. Phys. JETP 93, 356 (1987); Yu. Kagan, N.V. Prokof'ev, Sov. Phys. JETP 90, 2176 (1986)

3. B.L. Altshuler and A.G. Aronov, Electron – electron correlations in disordered conductors, in Modern Problems in Condensed Matter Systems, vol. 10, p.1 (North Holland, Amsterdam, 1985)

4. W. Kohn and J.M. Luttinger, Phys. Rev. Lett., 15, 524 (1965)

5. J. Kondo, Prog. Theor. Phys., 32, 37 (1964); C.M. Varma, Y. Yaflet, Phys. Rev. B 13, 2959 (1976); A.C. Hewson, The Kondo Problem in Heavy Fermions (Cambridge University Press, Cambridge, 1993); K.G. Wilson, Rev. Mod. Phys., 47, 773 (1975); P.W. Anderson, G. Yuval, D.R. Hamann, Phys. Rev. B 1, 4464 (1970); P. Nozieres, Jour. of Low Temp. Phys., 17, 31 (1974)

6. A.M. Tsvelik, P.B. Wiegman, Adv. in Phys., 32, 453 (1983); N. Andrei, F. Furuya, J.H. Loerenstein, Rev. Mod. Phys., 55, 331 (1983); P. Coleman, Phys. Rev. B 29, 3035 (1984)

7. D.M. Newns and M. Read, Adv. in Phys., 56, 799 (1987) ); P. Coleman, Phys. Rev. B 35, 5072 (1987); H. Keiter and N. Grewe in Valence Fluctuations in Solids, p. 451 (North Holland, New York, 1981); C.M. Varma, arXiv:cond-mat/0510019 (2005) and references therein

8. J.S. Kim, B. Andraka, G.R. Stewart, Phys. Rev. B 45, 12081 (1992)

9. M.Yu. Kagan and A.G. Aronov, Czechoslovac. Jour. of Phys., 46, 2061 (1996), Proc. of LT-21 Conference, Prague (1996)

10. A.L. Rakhmanov, K.I. Kugel, Ya.M. Blanter, M.Yu. Kagan, Phys. Rev. B 63, 174424 (2001); M.Yu. Kagan and K.I. Kugel, Sov. Phys. Uspekhi, 171, 577 (2001); M.Yu. Kagan, D.I. Khomskii, M.V. Mostovoy, Eur. Phys. Jour. B12, 217, 1999

11. K.I. Kugel, A.L. Rakhmanov, A.O. Sboychakov, Phys. Rev. B 76, 195113 (2007)

12. M.Yu. Kagan, A.V. Chubukov, JETP Lett., 47, 525 (1988); M.A. Baranov, A.V. Chubukov, M.Yu. Kagan, Int. Jour. Mod. Phys. B 6, 2471 (1992); M.A. Baranov, D.V. Efremov, M.S. Mar'enko, M.Yu. Kagan, Sov. Phys. JETP 90, 861 (2000)

13. M.Yu. Kagan, Phys. Lett. A 152, 303 (1991); M.A. Baranov, M.Yu. Kagan, Sov. Phys. JETP 102, 313 (1991)

14. J. Kanamori, Prog. Theor. Phys., 30, 275 (1963); J. Hubbard, Proc. Roy. Soc. A 276, 235 (1963)

15. V.M. Galitskii, Sov. Phys. JETP 34, 151 (1958)

16. E.M. Lifshitz, L.P. Pitaevskii, Statistical Physics, Part 2 (Nauka, Moscow, 1973)

17. P. Bloom, Phys. Rev. B 12, 125 (1975)

18. A.A. Abrikosov, L.P. Gor'kov, I.E. Dzyaloshinskii, Methods of Quantum Field Theory in Statistical Physics (Dover, New York, 1975)

19. M.Yu. Kagan, P.Woelfle, unpublished

20. M.Yu. Kagan, Superconductivity and Superfluidity in Fermi-systems with repulsive interaction, Habilitation Thesis (P.L. Kapitsa Institute for Physical Problems, Moscow, 1994)

21. P.W. Anderson, Phys. Rev. Lett., 64, 1839 (1990); P.W. Anderson, Phys. Rev. Lett., 654, 3306 (1990); P.W. Anderson, Phys. Rev. Lett., 66, 3226 (1991)

22. J.R. Engelbrecht and M. Randeria, Phys. Rev. Lett., 65, 1032 (1990); J.R. Engelbrecht and M. Randeria, Phys. Rev. Lett., 66, 3225 (1990); J.R. Engelbrecht and M. Randeria, Phys. Rev. B 45, 12419 (1992)

23. M.Yu. Kagan, N.V. Prokof'ev, unpublished

24. G. Iche and P. Nozieres, Physica A 91, 485 (1978); P. Nozieres and L.T. de Dominics, Phys. Rev., 178, 1097 (1969)

25. P.W. Anderson, Phys. Rev. Lett., 18, 1049 (1958); P.W. Anderson, Phys. Rev. B 164, 352 (1967)

26. R.M. White, P. Fulde, Phys. Rev. Lett., 47, 1540 (1981); G. Zwicknagl, A. Yaresko, P. Fulde, Phys. Rev. B 68, 052508 (2003); G. Zwicknagl, A. Yaresko, P. Fulde, Phys. Rev. B 65, 081103 (2002); M. Koga, W. Liu, M. Dolg, P. Fulde, Phys. Rev. B 57, 10648 (1998); P. Fulde, Electron Correlations in Molecules and Solids (Springer, Berlin, 1995).

27. Yu. Kagan, K.A. Kikoin, N.V. Prokof'ev, JETP Lett., 56, 221 (1992)

28. I.M. Lifshitz and A.M. Kosevich, Zh. Eksp. Theor. Phys., 29, 730 (1955); D. Schoenberg, Magnetic Oscillations in Metals (Mir, Moscow, 1986); L. Taillefer and G.G. Lonzarich, Phys. Rev. Lett., 60, 1570 (1988); A. Wasserman, M. Springford and A.C. Hewson, Jour. of Phys. Cond. Matt., 1, 2669 (1989)

29. T. Ito, H. Kumigashira, H.-D. Kim, T. Takahashi, N. Kimura, Y. Haga, E. Yamamoto, Y. Onuki, H. Harima, Phys. Rev. B 59, 8923 (1999); A.J. Arko, J.J. Joyce, A.B. Andrews, J.D. Thompson, J.L. Smith, E. Moshopoulou, Z. Fisk, A.A. Menovsky, P.C. Canfield, C.G. Olson, Physica B 230–232, 16 (1997)

30. A. Berton, J. Chaussy, B. Cornut, J. Flouquet, J. Odin, J. Peyrard, F. Holtzberg, Phys. Rev. B 23, 3504 (1981); G.R. Stewart, Rev. Mod. Phys., 56, 755 (1984); H.R. Ott, Progr. Low. Temp. Phys., 11, 215 (1987)

31. E.M. Lifshitz, L.P. Pitaevskii, Physical Kinetics (Nauka, Moscow, 1979); L.D. Landau, Phys. Zeitschr. Sow., 10, 154 (1936)

32. Y. Imry, Introduction to Mesoscopic Physics (Oxford University Press, Oxford, 2002); Y. Aharonov, D. Bohm, Phys. Rev., 115, 485 (1959)

33. H. Kusunose, S. Yotsuhashi and K. Miyake, Phys. Rev. B 62, 4403 (2000)

34. P. Farkašovský, Phys. Rev. B 77, 085110 (2008)

35. E. M. Levin, V.K. Pecharsky, K.A. Gschneidner, Jr., and G. J. Miller, Phys. Rev. B 64, 235103 (2001)

36. K. Murase, S. Ishida, S. Takaoka, T. Okumura, H. Fujiyasu, A. Ishida, M. Aoki, Surface Science 170, 486 (1986)

37. F.Steglich, J. Aarts, C. D. Bredl, G. Cordier, F. R. de Boer, W. Lieke, and U. Rauchschwalbe, 1982, in Superconductivity in *d*- and *f*-band metals (Kernforschungszentrum, Karlsruhe, 1982); M. Sigrist and K. Ueda, Rev. Mod. Phys., 63, 239 (1991)

38. Y. Maeno, T.M. Rice and M. Sigrist, Phys. Today 54, 42 (2001); T.M. Rice and M. Sigrist, J. Phys.: Cond. Matt., 7, L643 (1995)

39. D.J. Scalapino, E. Loh, Jr., J.E. Hirsch, Phys. Rev. B 34, 8190 (1986)

40. K. Miyake, S. Schmitt-Rink, C.M. Varma, Phys. Rev. B 34, 6554 (1986)

41. P. Coleman, E. Miranda, A. Tsvelik, Phys. Rev. Lett., 70, 2960 (1993); R. Flint, M. Dzero, P. Coleman, Nature Physics 4, 643 (2008) and references therein; P. Coleman, in Handbook of Magnetism and Advanced Magnetic Materials, vol. 1: Fundamentals and Theory (John Wiley and Sons, Hoboken, 2007)

42. K.I. Kugel, D.I. Khomskii, Sov. Phys. Uspekhi 136, 621 (1982)

43. S. Ishihara, J. Inoue, and S. Maekawa, Phys. Rev. B 55, 8280 (1997)

44. M.Yu. Kagan and T.M. Rice, J. Phys.: Cond. Matt., 6, 3771 (1994)

45. K.I. Kugel, A.L. Rakhmanov, A.O. Sboychakov, D.I. Khomskii, Phys. Rev. B 78, 115113 (2008)

46. P. Chandra, P. Coleman, J. A. Mydosh, V. Tripathi, Nature 417, 831 (2002); P. Chandra, P. Coleman, J.A. Mydosh, V. Tripathi, J. Phys.: Cond. Mat., 15, S1965 (2003)

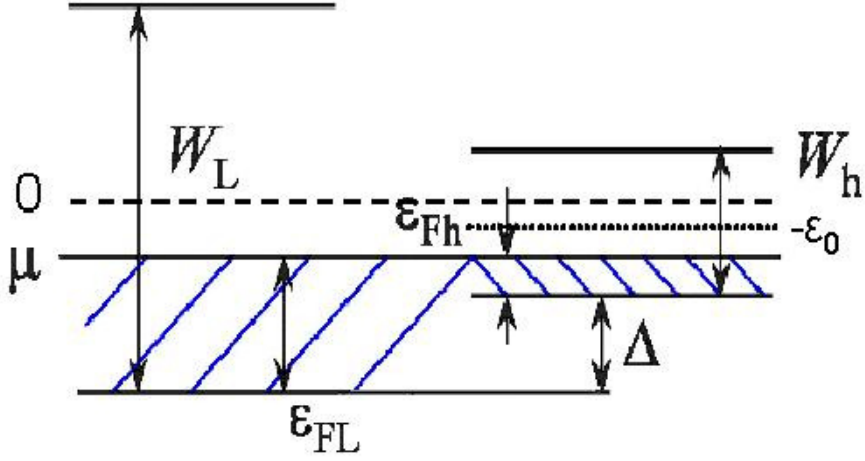

Fig.1. The band structure in the two-band model with one narrow band. $W_h$ and $W_L$ are the bandwidths of heavy and light electrons, $\varepsilon_{Fh}$ and $\varepsilon_{FL}$ are the Fermi energies $\Delta$ – is the energy difference between the bottoms of the heavy and light bands, $\Delta = -\varepsilon_0 + \dfrac{(W_L - W_h)}{2}$, where $(-\varepsilon_0)$ is the center of gravity of a heavy band. The center of gravity of a light band is at zero. $\mu$ – is chemical potential

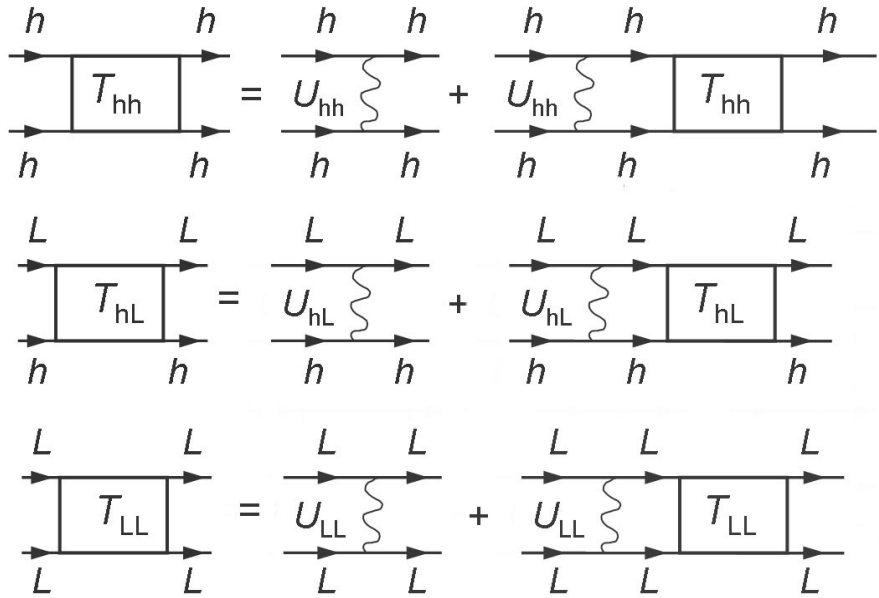

Fig.2. T-matrices $T_{hh}$, $T_{LL}$ and $T_{hL}$ for the two-band model with heavy (h) and light (L) electrons, $U_{hh}$ and $U_{LL}$ are the intraband Hubbard interactions, $U_{hL}$ is interband Hubbard interaction between heavy and light particles.

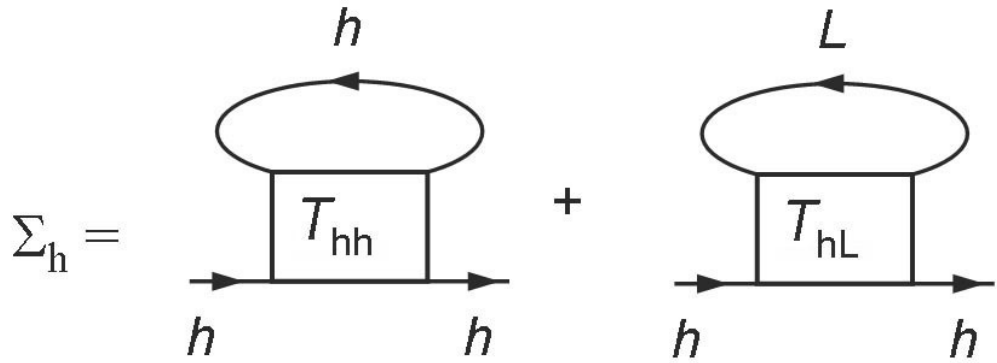

Fig.3 The T – matrix approximation for the self-energies of a heavy particle. $T_{hh}$ and $T_{hL}$ are the full T-matrices in substance. The diagrams for $\Sigma_L$ have the analogous character.

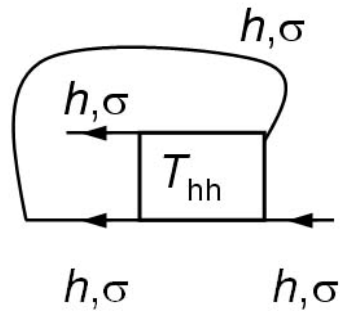

Fig.4. An exchange – type diagram for the self-energy $\Sigma_{hh}^{\sigma}$ which contains the matrix element $a^{+}_{\sigma}a^{+}_{\sigma}a_{\sigma}a_{\sigma}$ and thus is absent in the Hubbard model.

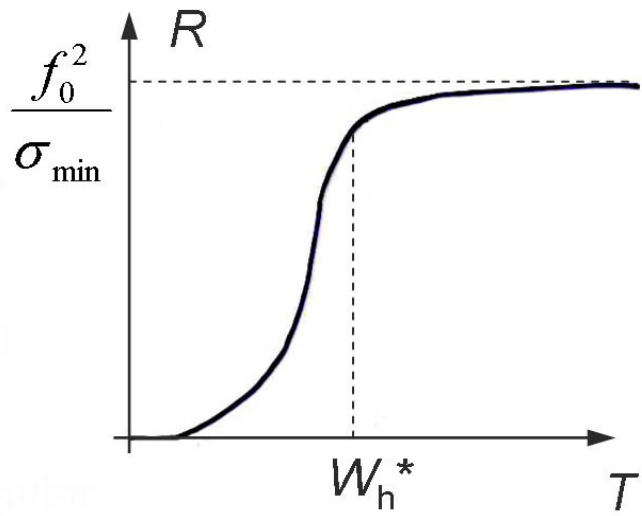

Fig.5. The resistivity characteristics $R(T)$ in the two-band model in 3D.

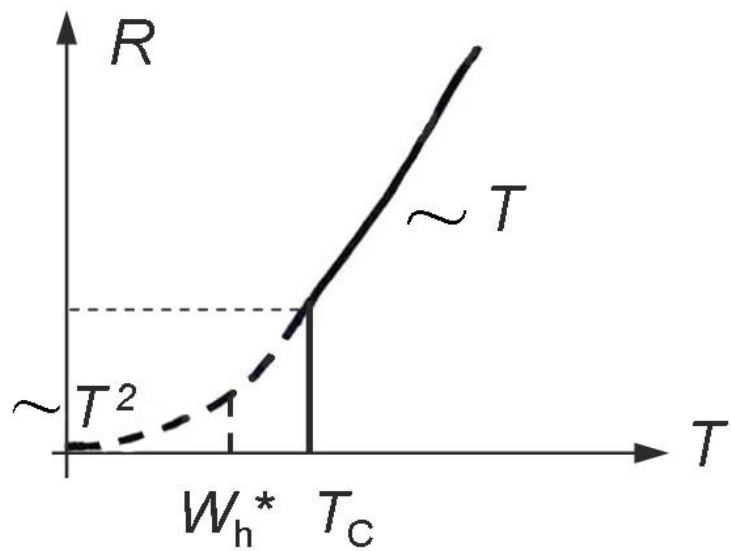

Fig.6. Resistivity $R(T)$ in superconducting material with a hidden heavy band for $W_h^* < T_C$ ($W_h^*$ is an effective width of a heavy band).

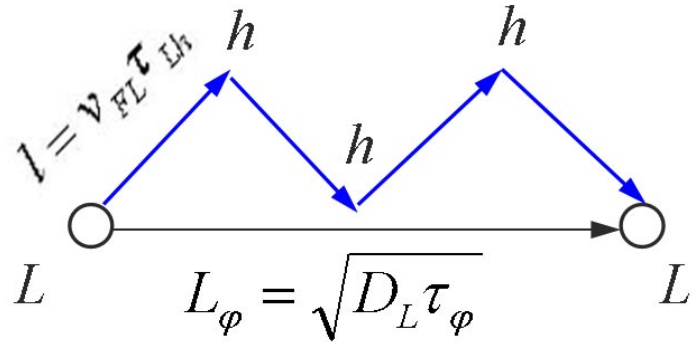

Fig.7. Multiple scattering of light particle on the heavy ones in between of the scattering of light particle on another light particle. $L_\varphi$ is a diffusive length, $l$ is elastic length, $D_L$ and $v_{FL}$ are diffusion coefficient and Fermi – velocity for light electrons, $\tau_{Lh}$ and $\tau_\varphi$ are elastic time for scattering of light electrons on heavy ones and inelastic (decoherence) time.

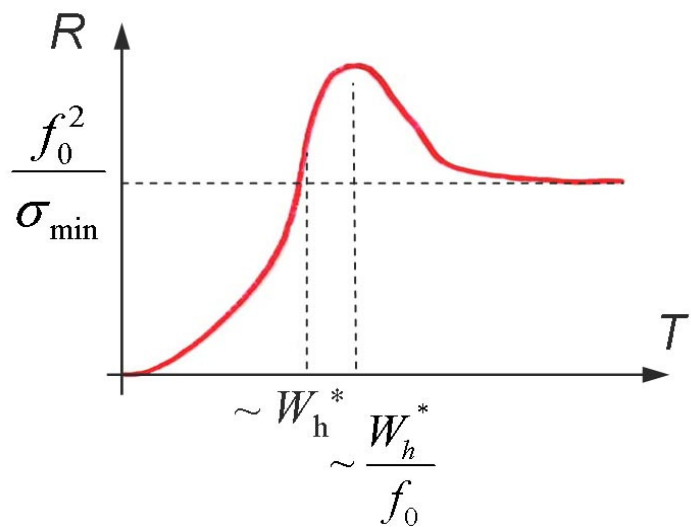

Fig.8. Resistivity $R(T)$ in a 2D case for the two-band model with one narrow band. It has a maximum and localization tail at higher temperatures $T > W_h^*$.

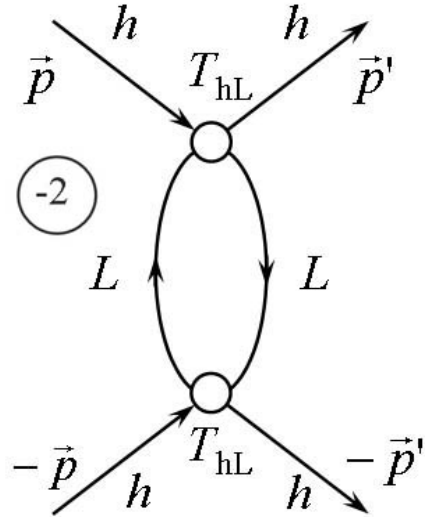

Fig.9. The leading contribution to the effective interaction $V_{eff}$ for the p-wave pairing of heavy particles via polarization of light particles. The open circles stand for the vacuum $T$ – matrix $T_{hL}$